\documentclass{article}

\usepackage{arxiv}

\usepackage[utf8]{inputenc} % allow utf-8 input
\usepackage[T1]{fontenc}    % use 8-bit T1 fonts
\usepackage{hyperref}       % hyperlinks
\usepackage{url}            % simple URL typesetting
\usepackage{booktabs}       % professional-quality tables
\usepackage{amsfonts}       % blackboard math symbols
\usepackage{nicefrac}       % compact symbols for 1/2, etc.
\usepackage{microtype}      % microtypography
\usepackage{lipsum}		% Can be removed after putting your text content
\usepackage{graphicx}
\usepackage{amsmath}
\usepackage{natbib}
\usepackage{framed}
\usepackage{doi}

\title{Foundations of ecological and evolutionary change}

\author{ \href{https://orcid.org/0000-0001-8343-4995}{\includegraphics[scale=0.06]{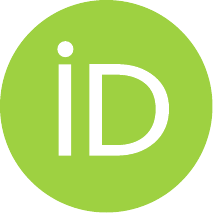}\hspace{1mm}A. Bradley Duthie} \\
	Department of Biological and Environmental Sciences \\
	University of Stirling \\
	Stirling, Scotland \\
	\texttt{alexander.duthie@stir.ac.uk} \\
	\And
	\href{https://orcid.org/0000-0002-2911-1105}{\includegraphics[scale=0.06]{orcid.pdf}\hspace{1mm}Victor J. Luque} \\
	Department of Philosophy\\
	University of Valencia\\
	Valencia, Spain \\
	\texttt{Victor.Luque@uv.es} \\
}

\date{}

\renewcommand{\headeright}{}
\renewcommand{\undertitle}{}
\renewcommand{\shorttitle}{Foundations of ecological and evolutionary change}

\hypersetup{
pdftitle={A template for the arxiv style},
pdfsubject={q-bio.NC, q-bio.QM},
pdfauthor={David S.~Hippocampus, Elias D.~Striatum},
pdfkeywords={First keyword, Second keyword, More},
}

\begin{document}
\maketitle

\begin{abstract}
Biological evolution is realised through the same mechanisms of birth
and death that underlie change in population density. The deep
interdependence between ecology and evolution is well-established, and
recent models focus on integrating eco-evolutionary dynamics to
demonstrate how ecological and evolutionary processes interact and feed
back upon each other. Nevertheless, a gap remains between the logical
foundations of ecology and evolution. Population ecology and evolution
have fundamental equations that define how the size of a population
(ecology) and the average characteristic within a population (evolution)
change over time. These fundamental equations are a complete and exact
description of change for any closed population, but how they are
formally linked remains unclear. We link the fundamental equations of
population ecology and evolution with an equation that sums how
individual characteristics interact with individual fitness in a
population. From this equation, we derive the fundamental equations of
population ecology and evolutionary biology (the Price equation). We
thereby identify an overlooked bridge between ecology and biological
evolution. Our unification formally recovers the equivalence between
mean population growth rate and evolutionary fitness and links this
change to ecosystem function. We outline how our framework can be used
to further develop eco-evolutionary theory.
\end{abstract}

\vspace{3mm}

% keywords can be removed

\section*{Introduction}

Theoretical unification is a powerful tool for scientific advancement.
Such unification has been a major goal in scientific research throughout
history (Smocovitis 1992; Kitcher 1993), and its value is perhaps most
evident in reconciling unconnected models and revealing new and
unexpected empirical predictions. In evolutionary biology, the Price
equation (Box 1) provides a unifying framework for evolutionary theory
by exhaustively and exactly describing evolutionary change for any
closed population (Price 1970; Luque 2017; Lehtonen et al. 2020). The
Price equation is therefore fundamental, in the sense that it binds
together all of evolutionary theory by formally defining what
evolutionary change is and is not (Price 1970; Rice 2004; Gardner 2008;
Frank 2017; Luque 2017; Luque and Baravalle 2021). Using this formal
definition, the scope of, and relationships among, sub-disciplines
within evolutionary theory can be clarified. For example, fundamental
equations of both population and quantitative genetics can be derived
from the Price equation (Queller 2017). This provides conceptual clarity
by demonstrating the logical consistency of different theoretical
frameworks within evolutionary biology. Our aim here is to propose an
equation that extends this conceptual clarity to include population size
change and thereby provide a formal and exact definition for joint
ecological and evolutionary change.

In biological populations, ecological change is caused by the same
processes of individual birth and death that cause evolutionary changes
in allele frequencies and phenotypes (Turchin 2001; Connor and Hartl
2004; Barfield et al. 2011). As with evolution, a fundamental equation
can exhaustively and exactly define population change. Unlike the Price
equation, this fundamental equation is perhaps self-evident. Population
change is simply the addition of individuals minus the removal of
individuals from current population size (\(N_{t}\)), which recovers the
new population size (\(N_{t+1}\); Box 2). By definition, the
relationship \(N_{t+1} = N_{t} + Births - Deaths\) applies to any closed
population. Turchin (2001) argues that general principles are needed to
establish a logical foundation for population ecology, and this simple
birth and death model and the consequences that logically follow from it
(e.g., exponential population growth) are fundamental to population
ecology. Any unifying definition of joint ecological and evolutionary
change must be able, when formalised, to derive both the Price equation
and this birth and death model.

\begin{framed}

\textbf{Box 1:} The Price equation is an abstract formula to represent
evolutionary change. Formulated originally in the early 1970s by George
Price (Price 1970, 1972), it postulates some basic properties that all
evolutionary systems must satisfy: change over time, ancestor and
descendant relations, and a character or phenotype (Rice 2004). Using
simple algebraic language, the Price equation represents evolutionary
change with the predominant notation,
\[\bar{w}\Delta\bar{z} = \mathrm{Cov}\left(w, z\right) + \mathrm{E}\left(w\Delta z\right).\]
In the above equation, \(\Delta\bar{z}\) is the change in the average
character value \(z\) over a time step of arbitrary length, \(w\) is an
individual's fitness, and \(\bar{w}\) average population fitness. On the
right-hand side of the equation, the first term is the covariance
between a character value \(z\) and fitness \(w\), which reflects
\(\bar{z}\) change attributable to differential survival and
reproduction. The second term is the expected value of \(w \Delta z\),
which reflects the extent to which offspring deviate from parents in
\(z\) (Rice 2004; Okasha 2006; Frank 2012). A more specific version of
the covariance term was already known within the quantitative and
population genetics tradition (Robertson 1966), usually representing the
action of natural selection. The Price equation adds an expectation term
and abstracts away from any specific mechanisms of replication or
reproduction, or mechanisms of inheritance. Its definitional nature and
lack of substantive biological assumptions has been portrayed both as a
strength (Rice 2004; Frank 2012; Luque 2017; Baravalle and Luque 2022),
and its greatest weakness. The abstract nature of the Price equation
places it at the top of the hierarchy of fundamental theorems of
evolution from which the rest (Robertson's theorem, Fisher's fundamental
theorem, breeder's equation, Hamilton's rule, adaptive dynamics, etc.)
can be derived (Lehtonen 2016, 2018, 2020; Queller 2017). This
abstractness is also key to developing a more general view of evolution
(Rice 2020; Luque and Baravalle 2021; Edelaar et al. 2023). In contrast,
some researchers consider the Price equation just a triviality (even
tautological), and useless without further modelling assumptions (van
Veelen 2005; van Veelen et al. 2012). The debate remains open (van
Veelen 2020; Baravalle et al. 2025).
    
\end{framed}

The union of ecological and evolutionary processes has long been
recognised (e.g., Darwin 1859; Fisher 1958; Pelletier et al. 2009), but
the rise of eco-evolutionary models, which incorporate both, is
relatively recent following a widespread recognition that ecology and
evolution can happen on similar timescales (Govaert et al. 2019;
Yamamichi et al. 2023). Currently, a universally recognised formal
definition of eco-evolutionary change is lacking, with some
theoreticians broadly interpreting ``eco-evolutionary dynamics'' to
allow for a separation of ecological and evolutionary timescales (Lion
et al. 2023) and others advocating for a more narrow interpretation in
which no such separation is permitted (Bassar et al. 2021). Bassar et
al. (2021) identify two types of eco-evolutionary models that follow
from these interpretations. The first type uses separate equations to
model population size change versus evolutionary change, thereby
allowing for any number of ecological, evolutionary, or environmental
feedbacks (e.g., Lion 2018; Patel et al. 2018; Lion et al. 2023). The
second type models population demographics as functions of quantitative
traits with ecological and evolutionary change following from
demographic processes and trait distributions operating on the same
timescale (e.g., Barfield et al. 2011; Simmonds et al. 2020; Jaggi et
al. 2024). Both model types can be very general, but like all predictive
models, they rely on simplifying assumptions for tractability (Levins
1966; Luque 2017). These simplifying assumptions are often grounded in
the Price equation to demonstrate accuracy and logical consistency when
modelling evolutionary change (e.g., Coulson and Tuljapurkar 2008;
Barfield et al. 2011; Rees and Ellner 2016; Lion 2018). For example,
Barfield et al. (2011) link their model back to Price (1970), which they
consider to be a ``universal law of evolution'', to place their
conclusions concerning stage structured evolution in the broader context
of evolutionary theory. The role of fundamental equations is therefore
important for unifying theory (Luque and Baravalle 2021), and we believe
that a fundamental equation of eco-evolutionary change has been
curiously overlooked.

We present an equation from which the fundamental equations of ecology
and evolutionary biology can be derived. Derivation follows by adding
assumptions that are specific to population ecology or evolution in the
same way that key equations of population genetics or quantitative
genetics can be derived from the Price equation by restricting the
domain of interest (e.g., to allele frequencies in the case of
population genetics, or to continuous phenotypes in the case of
quantitative genetics, Queller 2017). We propose our equation as a
formal definition of eco-evolutionary change.

\section*{A foundation for biological evolution and population
ecology}

To unify biological evolution and population ecology, we must reconcile
the Price equation (Box 1) with the general equation for population
change (Box 2). The Price equation is critical for partitioning
different components of biological change (Price 1970; Frank 1997;
Gardner 2008; Luque 2017; Queller 2017; Lehtonen 2018, 2020). It has
also been highly useful for integrating evolutionary theory across
disciplines (Fox 2006; Brantingham 2007; MacCallum et al. 2012; Frank
2015; Borgstede and Luque 2021; Godsoe et al. 2021; Ulrich et al. 2024).
These properties would seem to make it an intuitive starting point for a
logical foundation of ecology and evolution, perhaps through some kind
of mathematical equivalence (Page and Nowak 2002) or addition of terms
(Collins and Gardner 2009), or through the use of its recursive
structure (Kerr and Godfrey-Smith 2009; Frank 2012). But despite its
flexibility, the Price equation still relies on relative frequencies,
which must by definition sum to one (Frank 2015). This is because the
Price equation describes the average change in a population; the
frequency of entities is scaled thereby conserving total probability
(Frank 2015, 2016). But to recover the fundamental principle of
exponential population growth (Turchin 2001), this scaling must be
avoided.

\begin{framed}
\textbf{Box 2:} The number of individuals in any closed population
(\(N\)) at any given time (\(t + 1\)) is determined by the number of
individuals at \(t\) (\(N_{t}\)), plus the number of births (\(Births\))
minus the number of deaths (\(Deaths\)),
\[N_{t+1} = N_{t} + Births - Deaths.\] This equation is necessarily true
for any closed population. Despite its simplicity, it is a general
equation for defining population change and a starting point for
understanding population ecology. Turchin (2001) notes that a
consequence of this fundamental equation is the tendency for populations
to grow exponentially (technically geometrically in the above case where
time is discrete). This inherent underlying tendency towards exponential
growth persists even as the complexities of real populations, such as
structure, stochasticity, or density-dependent effects are added to
population models (Turchin 2001). Given the assumption that all
individuals in the population are identical, a per capita rate of birth,
\(Births_{t} = b_{t}N_{t}\), and death, \(Deaths_{t} = d_{t}N_{t}\), can be defined.
Rearranging and defining \(\lambda_{t} = 1 + b_{t} - d_{t}\) gives,
\(N_{t+1} = N_{t}\lambda_{t}\). Here \(\lambda_{t}\) is the finite rate of
increase (Gotelli 2001), and note that because \(0 \leq d_{t} \leq 1\),
\(\lambda_{t} \geq 0\). Verbally, the change in size of any closed
population equals its existing size times its finite rate of increase.
\end{framed}

We therefore begin with the most fundamental axioms underlying the
ecology and evolution of living systems (Rice 2004; Rice and
Papadopoulos 2009). In such systems, diversity is discontinuous, in the
sense that living systems are composed of discrete entities including
individual organisms and groups of organisms (Dobzhansky 1970). Our
framework is general enough that entities can be anything discrete, but
we will focus on each entity \(i\) as an individual organism. Change
occurs between the current time step \(t\) and a future time step
\(t + 1\), and time steps can be arbitrarily short or long. Let
\(\beta_{i}\) be the count of direct descendants of \(i\) at \(t + 1\)
(e.g., offspring if a time step is a single generation). Similarly, let
\(\delta_{i}\) be the count of deaths summed across \(i\) and any of its
descendants from \(t\) to \(t+1\). For example, if \(i\) and all of its
descendants persist at \(t + 1\), then \(\delta_{i} = 0\), or if a time
step is one season in an annual species, then \(\delta_{i} = 1\). All
individuals are defined by some characteristic \(z_{i}\), and
\(\Delta z_{i}\) defines any change in \(z_{i}\) from \(t\) to
\(t + 1\). Together, \(z_{i} + \Delta z_{i}\) is the average
characteristic across any individual and its descendants alive at
\(t+1\). The total number of individuals in the population is \(N\).
From this foundation, we define \(\Omega\) (which takes the same
measurement units as \(z\)) to be the sum of characteristic values in
\(t+1\),

\[\Omega = \sum_{i=1}^{N} \left(\beta_{i} - \delta_{i} + 1 \right)\left(z_{i} + \Delta z_{i} \right).
\tag{1}
\]

The foundation of eco-evolutionary change defined by eqn 1 is therefore
a statistical interaction between the demographic processes of birth and
death (\(\beta_{i} - \delta_{i} + 1\)) and individual characteristics
(\(z_{i} + \Delta z_{i}\)). From eqn 1, we can derive the most
fundamental equations of population ecology and evolutionary biology
through an appropriate interpretation of \(z\). Under more limited
interpretations of \(z\), we can also interpret \(\Omega\) as a metric
of ecosystem function.

\section*{Population ecology}

To recover the general equation for population ecology (Box 2), we
define \(z_{i}\) as the identity of \(i\) belonging to the population.
In other words, we set \(z_{i} = 1\) to indicate that \(i\) is one
member of the population and therefore contributes one unit to the total
population size. In this restricted case, \(z\) is count, which takes
the unit 1 (note that `individual' is a label, not a unit, see Newell
and Tiesinga 2019). Further, we assume that individual membership and
the unit contribution to population size does not change by setting
\(\Delta z_{i} = 0\). In this case,

\[\Omega = \sum_{i=1}^{N} \left(\beta_{i} - \delta_{i} + 1 \right).\]

We can now interpret \(\Omega\) as the population size at \(t+1\),
\(N_{t+1}\). Summing up \(\beta_{i}\), \(\delta_{i}\), and current
individuals tallies up the total number of individuals in the next time
step,

\[N_{t+1} = N_{t} + Births - Deaths.
\tag{2}
\]

Using the classical assumptions of population ecology (Box 2, Gotelli
2001), we can then recover the fundamental tendency for populations to
grow (or decline) exponentially (Turchin 2001).

\section*{Evolutionary biology}

Recovering the Price equation requires a few more steps. We start by
defining individual fitness,

\[w_{i} = \beta_{i} - \delta_{i} + 1.
\tag{3}
\]

In this definition, the longevity of the individual matters. All else
being equal, an individual that survives from \(t\) to \(t + 1\) has a
higher fitness than one that dies, even if both have the same
reproductive output. With this definition of fitness (eqn 3), we
substitute \(w_{i}\) into eqn 1,

\[\Omega = \sum_{i=1}^{N} \left(w_{i}z_{i} + w_{i}\Delta z_{i} \right).
\tag{4}
\]

We break eqn 4 down further and multiply each side by \(1/N\),

\[\frac{1}{N}\Omega = \frac{1}{N}\sum_{i=1}^{N} \left(w_{i}z_{i} \right) + \frac{1}{N}\sum_{i=1}^{N}\left( w_{i}\Delta z_{i} \right).
\tag{5}
\]

We rewrite the terms on the right-hand side of eqn 5 as expected values
and remove the subscripts,

\[\frac{1}{N}\Omega = \mathrm{E}\left(w z \right) + \mathrm{E}\left( w \Delta z  \right).
\tag{6}
\]

Now we must consider the total conservation of probability (Frank 2015,
2016). In eqn 6, \(\Omega\) is the total sum trait values (\(z_{i}\))
across the entire population at \(t + 1\) divided by the number of
individuals (\(N\)) in the population at \(t\). But the size of the
population can change from \(t\) to \(t + 1\). To recover mean trait
change for the Price equation (and therefore conserve total
probability), we need to account for this change in population size. We
cannot treat \(\Omega/N\) as the mean of \(z\) at \(t+1\) (\(\bar{z}'\))
because we need to weight \(N\) by the mean fitness of the population at
\(t\) to account for any change in population size from \(t\) to
\(t+1\). We need to multiply the mean trait value \(\bar{z}'\) (at
\(t + 1\)) by the mean fitness \(\bar{w}\) (at \(t\)) to recover the
mean contribution of the \(N\) individuals at \(t\) to the total
\(\Omega\) (Case and Taper 2000; Ewens 2014). Consequently,

\[\Omega = N\bar{w}\bar{z}'.
\tag{7}
\]

Equation 7 conserves the total probability and recovers \(\Omega\) as
the summed trait value, which has the same measurement units as \(z\)
and equals expected population growth at \(t\) times mean trait value at
\(t + 1\). This is consistent with the population ecology derivation
from the previous section where \(z_{i} = 1\) by definition, and
\(\Omega = N_{t+1}\). We can therefore rewrite eqn 6,

\[\bar{w}\bar{z}' = \mathrm{E}\left(w z \right) + \mathrm{E}\left( w \Delta z  \right).
\tag{8}
\]

We can rearrange eqn 8 to derive the Price equation by expressing
covariance as
\(\mathrm{Cov}(X,Y) = \mathrm{E}(XY) - \mathrm{E}(X)\mathrm{E}(Y)\), and
therefore
\(\mathrm{E}(XY) = \mathrm{Cov}(X,Y) + \mathrm{E}(X)\mathrm{E}(Y)\).
Substituting into eqn 8,

\[\bar{w}\bar{z}' = \mathrm{Cov}\left(w ,z \right) + \bar{w}\bar{z} + \mathrm{E}\left( w \Delta z  \right).\]

From here,

\[\bar{w}\left(\bar{z}' - \bar{z}\right) = \mathrm{Cov}\left(w ,z \right) + \mathrm{E}\left( w \Delta z  \right).\]

Since \(\Delta \bar{z} = \left(\bar{z}' - \bar{z}\right)\),

\[\bar{w}\Delta \bar{z} = \mathrm{Cov}\left(w ,z \right) + \mathrm{E}\left( w \Delta z  \right).
\tag{9}
\]

Equation 9 is the Price equation. From eqn 1, which describes
fundamental birth and death processes in a population, we can therefore
derive both the most fundamental model of population ecology (eqn 2; Box
2) and the fundamental equation of evolution (eqn 9; Box 1).

\section*{Ecosystem function}

In some cases, \(\Omega\) could also be interpreted as the total
contribution of a population to ecosystem function. This is restricted
to cases in which \(z\) is a characteristic defining an absolute
quantity measured at the whole organism level such as biomass, seed
production, carbon capture, flower visits, or nutrient consumption (Collins and Gardner 2009). In
such cases, the sum across individuals gives a meaningful total quantity
for the population (e.g., the total biomass or seeds produced in the
population). When \(z\) is instead defined by relative organism-level
measurements such wing loading, nutrient ratio, or diet composition, or
when \(z\) is measured at a level of biological organisation below the
organism (e.g., average cell volume or leaf surface area), \(\Omega\)
does not have a clear population-level interpretation. In such cases,
eqn 1 still defines eco-evolutionary change; the interpretation of
\(\Omega\) by itself is just not as interesting, biologically. Box 3
provides an instructive example of three plants with different fitnesses
and fruit masses.

\begin{framed}
\textbf{Box 3:} As an instructive example of our framework, consider a
population of \(N_{t} = 3\) annual plants in which individual total
fruit mass (kg) is measured, and change is observed over a year. For all
plants, \(\delta_{i} = 1\) (because annual plants die between \(t\) and
\(t+1\)), and let plant fecundities be \(\beta_{1} = 1\),
\(\beta_{2} = 1\), and \(\beta_{3} = 2\). Applying eqn 1 to population
change such that \(\Omega = N_{t+1}\), \(z_{i} = 1\) and
\(\Delta z_{i} = 0\),
\(N_{t+1} = (1 - 1 + 1)(1 + 0) + (1 - 1 + 1)(1 + 0) + (2 - 1 + 1)(1 + 0) = 4\)
(note \(\bar{w} = 4/3\), so \(N_{t}\bar{w} = 3(4/3) = 4\)). Focusing
next on the characteristic of total fruit mass, let \(z_{1} = 0.8\) kg,
\(z_{2} = 1.0\) kg, and \(z_{3} = 1.5\) kg. Also let
\(\Delta z_{i} = 0.1\) for all plants to reflect a change in soil
environment from \(t\) to \(t + 1\). In this case, population fruit
yield at \(t+1\) is
\(\Omega = (1 - 1 + 1)(0.8 + 0.1) + (1 - 1 + 1)(1.0 + 0.1) + (2 - 1 + 1)(1.5 + 0.1) = 5.2\)
kg. At \(t\), mean fruit yield per plant was
\((0.8 + 1.0 + 1.5)/3 = 1.1\) kg, but at \(t+1\), mean fruit yield per
plant is \([(0.8 + 0.1) + (1.0 + 0.1) + 2(1.5 + 0.1)]/4 = 1.3\) kg. Note
that
\(\mathrm{Cov}(w, z) = (1/N)\sum_{i}^{N}(z_{i} - \bar{z})(w_{i} - \bar{w}) = (1/3)\left[(0.8 - 1.1)(1-4/3) + (1 - 1.1)(1-4/3) + (1.5 - 1.1)(2-4/3) \right]  = 2/15\)
and \(\mathrm{E}(w \Delta z) = 4/3 \times 0.1 = 2/15\), so applying the
Price equation, \(\bar{w}\Delta \bar{z} = 2/15 + 2/15 = 4/15\). Since
\(\bar{w} = 4/3\), multiplying both sides of the equation by \(3/4\)
returns \(\Delta \bar{z} = 0.2\), which is the mean difference in fruit
yield between \(t+1\) and \(t\). The framework expressed in eqn 1
thereby links population change, evolutionary change, and ecosystem
function.
\end{framed}

\section*{Discussion}

An important aspect of scientific progress is the ability to connect
disparate theories and models to show how specific empirical and
theoretical models are logical (mathematical) consequences of more
fundamental ones (Nagel 1961; Morrison 2000). Rather than making
simplifying assumptions, as is the approach for specific ecological and
evolutionary models, we focus on fundamental axioms that are universal
to closed biological systems: discrete individuals, birth, death, and
change over time. We define an abstract sum (\(\Omega\)), to which all
individuals in the population contribute. From the basic axiom that each
individual is one member of a specific population (\(z_{i} = 1\) and
\(\Delta z_{i} = 0\)), we recover the most general equation of
population ecology (Box 2). By defining individual fitness (\(w_{i}\))
and applying the total conservation of probability to individual
frequencies (Frank 2015, 2016), we recover the most fundamental equation
of evolution, the Price equation (Box 1). Our eqn 1 thereby provides a
foundation for defining eco-evolutionary change in any population.

The Price equation provides a complete and exact description of
evolution in any closed evolving system (Price 1970; Frank 2012). It is
derived by rearranging the mathematical notation defining changes in the
frequencies and characteristics of any type of entity (e.g.,
individuals, alleles, Price 1970; Gardner 2008; Luque 2017). This
derivation partitions total characteristic change into different
components, making it possible to isolate evolutionary mechanisms (e.g.,
selection) and levels of biological organisation (e.g., group,
individual, Frank 1995, 2012; Kerr and Godfrey-Smith 2009; Luque 2017;
Okasha and Otsuka 2020). Because of its abstract nature and lack of any
system-specific assumptions, the Price equation is not dynamically
sufficient and makes no predictions about what will happen in any
particular system (Gardner 2020). Its role is not to predict, but to
formally and completely define and separate components of evolutionary
change (Baravalle et al. 2025). The same is true of the general equation
for population change (eqn 2), at least as we have used it here where it
serves to define what population change means in ecology. This equation
formally and completely describes population change in terms of births
and deaths. In eqn 1, we therefore have a fundamental equation from
which we can derive complete ecological and evolutionary change in any
closed biological population. Like all fundamental equations, our
equation is necessarily abstract and not dynamically sufficient. We
believe that it will be useful for eco-evolutionary theory in a similar
way that the Price equation is useful for evolutionary theory:
potentially facilitating specific model development and identifying new
conceptual insights, unresolved errors, and sources of model
disagreements (see below and Supporting Information S1).

Our unification recovers the equivalence between the finite rate of
increase \(\lambda\) (Box 2) and population mean evolutionary fitness
\(\bar{w}\) (Box 1). The population growth equation
\(N_{t+1} = N_{t}\lambda\) can always be rewritten as
\(N_{t+1} = N_{t}\bar{w}\). This specific equivalence has been proposed
before (e.g., Lande 1976), as has the broader relationship between
population growth rate and evolutionary fitness (e.g., Fisher 1930;
Charlesworth 1980; Lande 1982; Case and Taper 2000; Roff 2008; Lion
2018). We show this from first principles and clarify the relationship
between fitness and population growth. Over an arbitrary length of time,
fitness is properly defined as \(w_{i} = \beta_{i} - \delta_{i} + 1\).
Rates of change in ecology and evolution are reflected in the first and
second statistical moments of fitness, respectively. Population growth
rate reflects mean fitness \(\bar{w}\), while the rate of evolutionary
change reflects the variance in fitness \(\mathrm{Var}(w)/\bar{w}\)
(i.e., Fisher's fundamental theorem, Frank 1997; Rice 2004; Queller
2017).

Our unification may also help explain, at least partially, some of the
success of classical population genetic models. For decades, population
genetics (and to some extent quantitative genetics) has been accused of
being a reductionist view of evolution, reducing everything to changes
in allele frequencies and abstracting away from individuals and their
environments (the ecological interactions, MacColl 2011). This has been
a line of argumentation by some defenders of the so-called Extended
Synthesis (Pigliucci 2009), especially in relation to niche construction
(Odling-Smee et al. 2003). Famously, Mayr (1959) characterised
population genetics as a simple input and output of genes, analogous to
``the adding of certain beans to a beanbag and the withdrawing of
others'' (also called ``beanbag genetics''). Historical critics of
population genetics could not articulate a clear explanation for why it
works so well despite all of its idealisations and simplifications. From
the Price equation, we are able to recover classical population and
quantitative genetic models (Queller 2017) and develop new ones (Rice
2004, 2020; Luque 2017; Lion 2018). Our eqn 1 contains ecology at its
core, and we show how the Price equation logically follows from it after
accounting for absolute population growth (eqn 7). We therefore conclude
that population and quantitative genetic equations contain ecology (no
matter how hidden), and the ecological nature of evolution is implicit
in population and quantitative genetic models.

We have focused on the dynamics of a closed population, and in doing so
leave ecological and evolutionary change attributable to migration for
future work. In population ecology, immigration and emigration can be
incorporated by adding a term for each to the right-hand side of the
equation in Box 2 (Gotelli 2001). In evolution, because the Price
equation relies on mapping ancestor-descendant relationships, accounting
for migration is more challenging. Kerr and Godfrey-Smith (2009)
demonstrate how the Price equation can be extended to allow for
arbitrary links between ancestors and descendants, thereby extending the
Price equation to allow for immigration and emigration. Frank (2012)
presents a simplified version of Kerr and Godfrey-Smith (2009) that
allows some fraction of descendants to be unconnected to ancestors. In
both ecology and evolution, accounting for migration is done through the
use of additional terms on the right-hand side of the equations.

We believe our fundamental equation to be complete and exact for any
closed population. It therefore implicitly includes any effects of
density dependence on population growth (see Box 2), or any social
effects on evolutionary change (see Box 1). Both of these effects can be
made explicit by specifying how other individuals in a population affect
birth and death of a focal individual. We demonstrate this by deriving
more specific models of density dependent population growth and
multi-level selection in Supporting Information S1.

We have shown that we can derive the fundamental equations of population
ecology and biological evolution from a single unifying equation.
Lastly, we propose our eqn 1 as a potential starting point for defining
ecosystem function and further conceptual unification between ecology,
evolution, and ecosystem function. The Price equation has previously
been used to investigate ecosystem function (Loreau and Hector 2001; Fox
2006), but not with any attempt towards conceptual unification with
evolutionary biology. For example, Fox (2006) applied the abstract
properties of the Price equation to partition total change in ecosystem
function into separate components attributable to species richness,
species composition, and context dependent effects. This approach
provides a framework for comparing the effects of biodiversity on
ecosystem function in empirical systems (Fox 2006; Winfree et al. 2015;
Mateo-Tomás et al. 2017). Instead, our eqn 1 defines \(\Omega\) as total
ecosystem function contributed by a focal population. It is therefore
possible to investigate ecological, evolutionary, and ecosystem function
change from the same shared framework.

Over 120 years ago, Needham (1904) described natural history as ``the
study of the phenomena of fitness''. Fundamentally, we show why
eco-evolutionary change is a statistical interaction between fitness and
individual characteristics. From this definition, we can recover both
population size change and evolutionary change.

\section*{Acknowledgements}

This manuscript was supported by joint funding between the French
Foundation for Research on Biodiversity (FRB) Centre for the Synthesis
and Analysis of Biodiversity (CESAB) and the German Centre for
Integrative Biodiversity Research (sDiv). It was written as part of the
Unification of Modern Coexistence Theory and Price Equation (UNICOP)
project. Victor J. Luque was also supported by the Spanish Ministry of
Science and Innovation (Project: PID2021-128835NB-I00), and the
Conselleria d'Innovaci\'{o}, Universitats, Ci\`{e}ncia i
Societat Digital -- Generalitat Valenciana (Project: CIGE/2023/16). We
are grateful for many conversations with S\'{e}bastien Lion, Kelsey
Lyberger, Swati Patel, and especially Lynn Govaert, whose questions
helped us clarify the relationship between population growth and
fitness. Brad Duthie would also like to thank Matt Tinsley, Brent
Danielson, and Stan Harpole. Victor J. Luque would also like to thank
Lorenzo Baravalle, Pau Carazo, Santiago Ginnobili, Manuel Serra, and
Ariel Roff\'{e}. Lastly, we thank two anonymous reviewers and Andy
Gardner for helpful comments.

\section*{Author Contributions}

Both authors came up with the question idea. Following many discussions
between the authors, ABD proposed the initial equation with subsequent
exploration and development from both authors. Both authors contributed
to the writing.

\section*{Competing Interests}

The authors declare no competing interests.

\section*{Data Availability}

This work does not include any data.

\section*{References}

\hangindent=0.5cm
Baravalle, L., and V. J. Luque. 2022.
\href{https://doi.org/10.1387/theoria.21940}{Towards a Pricean
foundation for cultural evolutionary theory}. Theoria 37:209--231.

\hangindent=0.5cm
Baravalle, L., A. Roff\'{e}, V. J. Luque, and S. Ginnobili. 2025.
\href{https://doi.org/10.1007/s13752-024-00482-4}{The value of Price}.
Pages 12--24 \emph{in} Biological Theory (Vol. 20).

\hangindent=0.5cm
Barfield, M., R. D. Holt, and R. Gomulkiewicz. 2011.
\href{https://doi.org/10.1086/658903}{Evolution in stage-structured
populations}. American Naturalist 177:397--409.

\hangindent=0.5cm
Bassar, R. D., T. Coulson, J. Travis, and D. N. Reznick. 2021.
\href{https://doi.org/10.1111/ele.13712}{Towards a more precise -- and
accurate -- view of eco-evolution}. Ecology Letters 24:623--625.

\hangindent=0.5cm
Borgstede, M., and V. Luque. 2021. The covariance based law of effect.
Behavior and Philosophy 49:63--81.

\hangindent=0.5cm
Brantingham, P. J. 2007. \href{https://doi.org/10.2307/40035853}{{A
unified evolutionary model of archaeological style and function based on
the Price equation}}. American Antiquity 72:395--416.

\hangindent=0.5cm
Case, T. J., and M. L. Taper. 2000. Interspecific competition,
environmental gradients, gene flow, and the coevolution of species'
borders. American Naturalist 155:583--605.

\hangindent=0.5cm
Charlesworth, B. 1980. Evolution in age-structured populations.
Cambridge Studies in Mathematical Biology. Cambridge University Press,
Cambridge.

\hangindent=0.5cm
Collins, S., and A. Gardner. 2009.
\href{https://doi.org/10.1111/j.1461-0248.2009.01340.x}{{Integrating
physiological, ecological and evolutionary change: A Price equation
approach}}. Ecology Letters 12:744--757.

\hangindent=0.5cm
Connor, J., and D. L. Hartl. 2004. {A Premier of Ecological Genetics}.
Sinauer Associates Incorporated.

\hangindent=0.5cm
Coulson, T., and S. Tuljapurkar. 2008.
\href{https://doi.org/10.1086/591693}{The dynamics of a quantitative
trait in an age-structured population living in a variable environment}.
American Naturalist 172:599--612.

\hangindent=0.5cm
Darwin, C. 1859. The Origin of Species. Penguin.

\hangindent=0.5cm
Dobzhansky, T. 1970. Genetics of the Evolutionary Process (Vol. 139).
Columbia University Press.

\hangindent=0.5cm
Edelaar, Pim, J. Otsuka, and V. J. Luque. 2023.
\href{https://doi.org/10.1111/brv.12910}{{A generalised approach to the
study and understanding of adaptive evolution}}. Biological Reviews
98:352--375.

\hangindent=0.5cm
Ewens, W. J. 2014.
\href{https://doi.org/10.1007/s10539-013-9412-0}{Grafen, the Price
equations, fitness maximization, optimisation and the fundamental
theorem of natural selection}. Biology and Philosophy 29:197--205.

\hangindent=0.5cm
Fisher, R. A. 1930. The Genetical Theory of Natural Selection. Oxford
University Press, Oxford, UK.

\hangindent=0.5cm
Fisher, R. A. 1958. The Genetical Theory of Natural Selection (2nd ed.).
Dover.

\hangindent=0.5cm
Fox, J. W. 2006.
\href{https://doi.org/10.1890/0012-9658(2006)87\%5B2687:utpetp\%5D2.0.co;2}{{Using
the Price equation to partition the effects of biodiversity loss on
ecosystem function}}. Ecology 87:2687--2696.

\hangindent=0.5cm
Frank, S. A. 1995. \href{https://doi.org/10.1006/jtbi.1995.0148}{{George
Price's contributions to evolutionary genetics}}. Journal of Theoretical
Biology 175:373--388.

\hangindent=0.5cm
---------. 1997.
\href{https://doi.org/10.1111/j.1558-5646.1997.tb05096.x}{{The Price
equation, Fisher's fundamental theorem, kin selection, and causal
analysis}}. Evolution 51:1712--1729.

\hangindent=0.5cm
---------. 2012.
\href{https://doi.org/10.1111/j.1420-9101.2012.02498.x}{{Natural
selection. IV. The Price equation}}. Journal of Evolutionary Biology
25:1002--1019.

\hangindent=0.5cm
---------. 2015. \href{https://doi.org/10.3390/e17107087}{{D'Alembert's
direct and inertial forces acting on populations: The Price equation and
the fundamental theorem of natural selection}}. Entropy 17:7087--7100.

\hangindent=0.5cm
---------. 2016. \href{https://doi.org/10.3390/e18050192}{{Common
probability patterns arise from simple invariances}}. Entropy 18:1--22.

\hangindent=0.5cm
---------. 2017. \href{https://doi.org/10.1002/ece3.2922}{{Universal
expressions of population change by the Price equation: Natural
selection, information, and maximum entropy production}}. Ecology and
Evolution 1--16.

\hangindent=0.5cm
Gardner, A. 2008. \href{https://doi.org/10.1016/j.cub.2008.01.005}{{The
Price equation}}. Current Biology 18:198--202.

\hangindent=0.5cm
---------. 2020. \href{https://doi.org/10.1098/rstb.2019.0361}{{Price's
equation made clear}}. Philosophical Transactions of the Royal Society
B: Biological Sciences 375:20190361.

\hangindent=0.5cm
Godsoe, W., K. E. Eisen, D. Stanton, and K. M. Sirianni. 2021.
\href{https://doi.org/10.1007/s12080-020-00478-3}{Selection and
biodiversity change}. Theoretical Ecology 14:367--379.

\hangindent=0.5cm
Gotelli, N. J. 2001. A Primer of Ecology. Sinauer associate. Inc.
Sunderland, MA.

\hangindent=0.5cm
Govaert, L., E. A. Fronhofer, S. Lion, C. Eizaguirre, D. Bonte, M. Egas,
A. P. Hendry, et al. 2019.
\href{https://doi.org/10.1111/1365-2435.13241}{{Eco-evolutionary
feedbacks---Theoretical models and perspectives}}. Functional Ecology
33:13--30.

\hangindent=0.5cm
Jaggi, H., W. Zuo, R. Kentie, J. M. Gaillard, T. Coulson, and S.
Tuljapurkar. 2024. \href{https://doi.org/10.1111/ele.14551}{Density
dependence shapes life-history trade-offs in a food-limited population}.
Ecology Letters 27:e14551.

\hangindent=0.5cm
Kerr, B., and P. Godfrey-Smith. 2009.
\href{https://doi.org/10.1111/j.1558-5646.2008.00570.x}{{Generalization
of the Price equation for evolutionary change}}. Evolution 63:531--536.

\hangindent=0.5cm
Kitcher, P. 1993. {The Advancement of Science}. Oxford University Press,
New York.

\hangindent=0.5cm
Lande, R. 1976. Natural selection and random genetic drift in phenotypic
evolution. Evolution 30:314--334.

\hangindent=0.5cm
---------. 1982. \href{https://doi.org/10.2307/1936778}{{A quantitative
genetic theory of life history evolution}}. Ecology 63:607--615.

\hangindent=0.5cm
Lehtonen, J. 2016.
\href{https://doi.org/10.1016/j.tree.2016.07.006}{{Multilevel selection
in kin selection language}}. Trends in Ecology and Evolution 31:752--762.

\hangindent=0.5cm
---------. 2018. \href{https://doi.org/10.1086/694891}{{The Price
equation, gradient dynamics, and continuous trait game theory}}.
American Naturalist 191:146--153.

\hangindent=0.5cm
---------. 2020. \href{https://doi.org/10.1098/rstb.2019.0362}{The Price
equation and the unity of social evolution theory}. Philosophical
Transactions of the Royal Society B: Biological Sciences 375:20190362.

\hangindent=0.5cm
Lehtonen, J., S. Okasha, and H. Helanter\"{a}. 2020.
\href{https://doi.org/10.1098/rstb.2019.0350}{{Fifty years of the Price
equation}}. Philosophical Transactions of the Royal Society B:
Biological Sciences 375:20190350.

\hangindent=0.5cm
Levins, R. 1966. \href{https://doi.org/10.2307/27836590}{The strategy of
model building in population biology}. American Naturalist 54: 421-431.

\hangindent=0.5cm
Lion, S. 2018. \href{https://doi.org/10.1086/694865}{{Theoretical
approaches in evolutionary ecology: environmental feedback as a unifying
perspective}}. American Naturalist 191:21--44.

\hangindent=0.5cm
Lion, S., A. Sasaki, and M. Boots. 2023.
\href{https://doi.org/10.1111/ele.14183}{Extending eco-evolutionary
theory with oligomorphic dynamics}. Ecology Letters 26:S22--S46.

\hangindent=0.5cm
Loreau, M., and A. Hector. 2001.
\href{https://doi.org/10.1038/35083573}{{Partitioning selection and
complementarity in biodiversity experiments}}. Nature 412:72--76.

\hangindent=0.5cm
Luque, V. J. 2017. \href{https://doi.org/10.1007/s10539-016-9538-y}{{One
equation to rule them all: a philosophical analysis of the Price
equation}}. Biology and Philosophy 32:1--29.

\hangindent=0.5cm
Luque, V. J., and L. Baravalle. 2021.
\href{https://doi.org/10.1007/s11229-021-03339-6}{{The mirror of
physics: on how the price equation can unify evolutionary biology}}.
Synthese 199:12439--12462.

\hangindent=0.5cm
MacCallum, R. M., M. Mauch, A. Burt, and A. M. Leroi. 2012.
\href{https://doi.org/10.5061/dryad.h0228}{{Evolution of music by public
choice}}. Proceedings of the National Academy of Sciences
109:12081--12086.

\hangindent=0.5cm
MacColl, A. D. C. 2011.
\href{https://doi.org/10.1016/j.tree.2011.06.009}{{The ecological causes
of evolution}}. Trends in Ecology and Evolution 26:514--522.

\hangindent=0.5cm
Mateo-Tom\'{a}s, P., P. P. Olea, M. Mole\'{o}n, N. Selva, and J. A.
S\'{a}nchez-Zapata. 2017. \href{https://doi.org/10.1111/geb.12673}{{Both
rare and common species support ecosystem services in scavenger
communities}}. Global Ecology and Biogeography 26:1459--1470.

\hangindent=0.5cm
Mayr, E. 1959. \href{https://doi.org/10.1101/SQB.1959.024.01.003}{Where
are we? Genetics and twentieth century Darwinism}. Pages 1--14
\emph{in} Cold Spring Harbor Symposia on Quantitative Biology (Vol. 24).

\hangindent=0.5cm
Morrison, M. 2000. Unifying Scientific Theories: Physical concepts and
mathematical structures. Cambridge University Press.

\hangindent=0.5cm
Nagel, E. 1961. The Structure of Science: Problems in the logic of
scientific explanation. Harcourt, Brace \& World, New York, NY, USA.

\hangindent=0.5cm
Needham, J. G. 1904. Is the course for college entrance
requirements best for those who go no further? Science 19:650--656.

\hangindent=0.5cm
Newell, D. B., and E. Tiesinga. 2019.
\href{https://doi.org/10.6028/NIST.SP.330-2019}{The International System
of Units (SI)}. NIST Special Publication 330.

\hangindent=0.5cm
Odling-Smee, F. J., K. N. Lala, and M. W. Feldman. 2003. Niche
Construction: The neglected process in evolution. Princeton University
Press.

\hangindent=0.5cm
Okasha, S. 2006.
\href{https://doi.org/10.1093/acprof:oso/9780199267972.001.0001}{Evolution
and the Levels of Selection}. Oxford University Press.

\hangindent=0.5cm
Okasha, S., and J. Otsuka. 2020.
\href{https://doi.org/10.1098/rstb.2019.0365}{{The Price equation and
the causal analysis of evolutionary change}}. Philosophical Transactions
of the Royal Society B: Biological Sciences 375:20190365.

\hangindent=0.5cm
Page, K. M., and M. A. Nowak. 2002.
\href{https://doi.org/10.1006/jtbi.2002.3112}{Unifying evolutionary
dynamics}. Journal of Theoretical Biology 219:93--98.

\hangindent=0.5cm
Patel, S., M. H. Cortez, and S. J. Schreiber. 2018.
\href{https://doi.org/10.1101/104505}{{Partitioning the effects of
eco-evolutionary feedbacks on community stability}}. American Naturalist
191:1--29.

\hangindent=0.5cm
Pelletier, F., D. Garant, and A. P. Hendry. 2009.
\href{https://doi.org/10.1098/rstb.2009.0027}{Eco-evolutionary
dynamics}. Philosophical Transactions of the Royal Society B: Biological Sciences
364:1483--1489.

\hangindent=0.5cm
Pigliucci, M. 2009.
\href{https://doi.org/10.1111/j.1749-6632.2009.04578.x}{{An extended
synthesis for evolutionary biology}}. Annals of the New York Academy of
Sciences 1168:218--228.

\hangindent=0.5cm
Price, G. R. 1970. \href{https://doi.org/10.1038/227520a0}{{Selection
and covariance}}. Nature 227:520-–521.

\hangindent=0.5cm
---------. 1972.
\href{https://doi.org/10.1111/j.1469-1809.1957.tb01874.x}{{Extension of
covariance selection mathematics}}. Annals of Human Genetics
35:485--490.

\hangindent=0.5cm
Queller, D. C. 2017. \href{https://doi.org/10.1086/690937}{{Fundamental
theorems of evolution}}. American Naturalist 189:345--353.

\hangindent=0.5cm
Rees, M., and S. P. Ellner. 2016.
\href{https://doi.org/10.1111/2041-210X.12487}{Evolving integral
projection models: Evolutionary demography meets eco-evolutionary
dynamics}. Methods in Ecology and Evolution 7:157--170.

\hangindent=0.5cm
Rice, S. H. 2004. {Evolutionary Theory: Mathematical and conceptual
foundations}. Sinauer Associates Incorporated.

\hangindent=0.5cm
---------. 2020.
\href{https://doi.org/10.1098/rstb.2019.0353}{{Universal rules for the
interaction of selection and transmission in evolution}}. Philosophical
Transactions of the Royal Society B: Biological Sciences 375:20190353.

\hangindent=0.5cm
Rice, S. H., and A. Papadopoulos. 2009.
\href{https://doi.org/10.1371/journal.pone.0007130}{{Evolution with
stochastic fitness and stochastic migration}}. PLoS One 4:e7130.

\hangindent=0.5cm
Robertson, A. 1966. \href{https://doi.org/10.1017/S0003356100037752}{{A
mathematical model of the culling process in dairy cattle}}. Animal
Science 8:95--108.

\hangindent=0.5cm
Roff, D. A. 2008.
\href{https://doi.org/10.1007/s12041-008-0056-9}{{Defining fitness in
evolutionary models}}. Journal of Genetics 87:339--348.

\hangindent=0.5cm
Simmonds, E. G., E. F. Cole, B. C. Sheldon, and T. Coulson. 2020.
\href{https://doi.org/10.1111/ele.13603}{Phenological asynchrony: A
ticking time-bomb for seemingly stable populations?} Ecology Letters 23:1766--1775.

\hangindent=0.5cm
Smocovitis, V. B. 1992. Unifying biology: The evolutionary
synthesis and evolutionary biology. Source: Journal of the History of
Biology 25:1--65.

\hangindent=0.5cm
Turchin, P. 2001.
\href{https://doi.org/10.1034/j.1600-0706.2001.11310.x}{Does population
ecology have general laws?} Oikos 94:17--26.

\hangindent=0.5cm
Ulrich, W., N. J. Gotelli, G. Strona, and W. Godsoe. 2024.
\href{https://doi.org/10.1016/j.ecolmodel.2024.110695}{Reconsidering the
Price equation: Benchmarking the analytical power of additive
partitioning in ecology}. Ecological Modelling 491:110695.

\hangindent=0.5cm
van Veelen, M. 2005.
\href{https://doi.org/10.1016/j.jtbi.2005.04.026}{{On the use of the
Price equation}}. Journal of Theoretical Biology 237:412--426.

\hangindent=0.5cm
---------. 2020. \href{https://doi.org/10.1098/rstb.2019.0355}{{The
problem with the Price equation}}. Philosophical Transactions of the
Royal Society B: Biological Sciences 375:20190355.

\hangindent=0.5cm
van Veelen, M., J. Garc\'{i}a, M. W. Sabelis, and M. Egas. 2012.
\href{https://doi.org/10.1016/j.jtbi.2011.07.025}{{Group selection and
inclusive fitness are not equivalent; the Price equation vs. models and
statistics}}. Journal of Theoretical Biology 299:64--80.

\hangindent=0.5cm
Winfree, R., J. W. Fox, N. M. Williams, J. R. Reilly, and D. P.
Cariveau. 2015. \href{https://doi.org/10.1111/ele.12424}{{Abundance of
common species, not species richness, drives delivery of a real-world
ecosystem service}}. Ecology Letters 18:626--635.

\hangindent=0.5cm
Yamamichi, M., S. P. Ellner, and N. G. Hairston. 2023.
\href{https://doi.org/10.1111/ele.14197}{Beyond simple adaptation:
Incorporating other evolutionary processes and concepts into
eco-evolutionary dynamics}. Ecology Letters 26:S16--S21.

\clearpage

\section*{Supporting Information S1}

This supporting information demonstrates how to derive well-established
models and develop new theory in population ecology and evolutionary
biology from equation 1 in the main text,

\[\Omega = \sum_{i=1}^{N} \left(\beta_{i} - \delta_{i} + 1 \right)\left(z_{i} + \Delta z_{i} \right).
\tag{1}
\]

In the main text, we derived both the Price equation and the birth-death
model from the above. Here we (1) integrate interactions between
individuals to recover density-dependent population growth, (2)
integrate groups within the population to recover multi-level selection,
and (3) integrate both to model a system in which multi-level selection
and density-dependent population change occur simultaneously. Finally,
we (4) expand on our derivation of multi-level selection to demonstrate
how it might be used to recover both community and population-level
evolutionary change, partitioning this change into different components.

\section*{1. Density-dependent population growth}

There are two potential ways to model the incorporation of density
dependence into population growth. We start with what is likely to be
the most familiar model to readers, which focuses on individual growth
rate \(r_{i}\). We then use a slightly different model focusing on
fitness \(w_{i}\). First, note that here we set \(\Omega = N_{t+1}\),
and \(z_{i} = 1\) and \(\Delta z_{i} = 0\) for all individuals as in the
main text. We can define \(r_{i} = \beta_{i} - \delta_{i}\) as the
individual growth rate for \(i\) (Lion 2018; Lion et al. 2023). In this
case,

\[N_{t+1} = \sum_{i=1}^{N_{t}}\left(r_{i} + 1\right)
\tag{S1.1}
\]

Mathematically, the most general approach here would be to define
individual growth as a function of the entire system \(\mathbf{E}\),
\(r_{i}(\mathbf{E})\), where \(\mathbf{E}\) is a vector with elements
including any parameters potentially relevant to \(r_{i}\). Taking this
approach would recover a version of eqn S1.1 in Lion (2018) and permit
any relationship between the system and a focal individual's growth.
Limiting our focus to the effects of other individuals (\(j\)) and
assuming that the effects of these individuals are additive, let
\(a_{ij|\cdot}\) be the effect of individual \(j\) on the growth rate
attributable to \(i\) conditioned on all other individuals within the
population such that
\(r_{i}\left(1 - \sum_{i = j}^{N}a_{ij|\cdot} \right)\) defines the
realised growth rate of \(i\),

\[N_{t+1} = \sum_{i=1}^{N_{t}}\left(r_{i}\left(1 - \sum_{i = j}^{N_{t}}a_{ij|\cdot} \right) + 1\right).
\tag{S1.2}
\]

Assuming that individual effects of \(j\) on \(i\) are also independent,
we can remove the condition in S1.2,

\[N_{t+1} = \sum_{i=1}^{N_{t}}\left(r_{i}\left(1 - \sum_{i = j}^{N_{t}}a_{ij} \right) + 1\right).\]

Further assuming that all individuals have the same per capita effect
such that \(a = a_{ij}\) for any \(i\) and \(j\) pair (as might be
reasonable given resource competition in a well-mixed population),

\[N_{t+1} = \sum_{i=1}^{N_{t}}\left(r_{i}\left(1 - a N_{t} \right) + 1\right).\]

If \(r_{i}\) values are identical,

\[N_{t+1} = N_{t} + r N_{t}\left(1 - a N_{t} \right).
\tag{S1.3}
\]

Equation S1.3 therefore recovers a classic version of a discrete time
logistic growth by making assumptions from an exact model of
eco-evolutionary change.

An alternative approach would be to model the effects of an individual
\(j\) on the fitness of \(i\) (\(w_{i}\)), thereby replacing eqn S1.1
with \(N_{t+1} = \sum_{i=1}^{N_{t}}w_{i}\) and replacing eqn S1.2 with,

\[N_{t+1} = \sum_{i=1}^{N_{t}}w_{i}\left(1 - \sum_{i = j}^{N_{t}}\alpha_{ij|\cdot} \right).
\]

Note that we have used \(\alpha_{ij}\) to represent the effect of \(j\)
on the fitness of \(i\) for clarity in the sections below. By making the
same assumptions of additivity, independence, and identical effects such
that \(\alpha = \alpha_{ij}\) for all \(j\) on \(i\), and assuming
fitness is equal (\(w_{i} = \bar{w}\)), we can derive,

\[N_{t+1} = \bar{w}N_{t}(1 - \alpha N_{t}).
\tag{S1.4}
\]

Hence, S1.4 is an alternative way to express logistic growth.

\section*{2. Multi-level selection}

We can recover multi-level selection from our eqn 1. Here we derive the
original form of the multi-level Price (1972) equation as it appears in
eqn 3.1 of Lehtonen (2020). Individuals belong to one of \(K\) total
groups where \(j\) indexes \(K\) groups and \(i\) indexes individuals.
Individuals do not overlap in group membership. The size of group \(j\)
is denoted as \(N_{j}\). Equation S2.1 below uses summations to
partition how individuals within each group contribute to \(\Omega\),

\[\Omega = \sum_{j=1}^{K}\sum_{i=1}^{N_{j}}\left(\beta_{ji} - \delta_{ji} + 1 \right)\left(z_{ji} + \Delta z_{ji}\right).
\tag{S2.1}
\]

In S2.1, indices \(\beta_{ji}\), \(\delta_{ji}\), and \(z_{ji}\)
identify individual \(i\) in group \(j\). We set
\(w_{ji} = \beta_{ji} - \delta_{ji} + 1\), and for simplicity let
\(\Delta z_{ji} = 0\) (i.e., no transmission bias, but see section 4
below),

\[\Omega = \sum_{j=1}^{K}\sum_{i=1}^{N_{j}}w_{ji}z_{ji}.\]

For ease of presentation, with no loss of generality, we assume all
group sizes are equal with a group size of \(N_{j} = n\) for all \(j\).
If group sizes differ, then weighted expectations and covariances are
instead needed (Lehtonen 2020). Given equal group sizes, the total
number of individuals (\(N\)) equals \(K \times n\), and,

\[\frac{\Omega}{K n} = \left(\frac{1}{K}\right)\left(\frac{1}{n}\right)\sum_{j=1}^{K}\sum_{i=1}^{n}w_{ji}z_{ji}.\]

Rearranging,

\[\frac{\Omega}{K n} = \frac{1}{K}\sum_{j=1}^{K}\frac{1}{n}\sum_{i=1}^{n}w_{ji}z_{ji}.\]

The inner summation can be rewritten as an expectation for group \(j\),

\[\frac{\Omega}{K n} = \frac{1}{K}\sum_{j=1}^{K}\mathrm{E}_{j}\left(w_{ji} z_{ji}\right).\]

As in the main text, we note
\(\mathrm{E}(XY) = \mathrm{Cov}(X,Y) + \mathrm{E}(X)\mathrm{E}(Y)\).
Defining \(Cov_{j}(w_{j}, z_{j})\) as the covariance between \(w_{ji}\)
and \(z_{ji}\) for group \(j\),

\[\frac{\Omega}{K n} = \frac{1}{K}\sum_{j=1}^{K}\left(\mathrm{Cov}_{j}\left(w_{ji}, z_{ji}\right) + \mathrm{E}_{j}\left(w_{ji}\right)\mathrm{E}_{j}\left(z_{ji}\right)\right).\]

We can separate the summation for each term,

\[\frac{\Omega}{K n} = \frac{1}{K}\sum_{j=1}^{K}\left(\mathrm{Cov}_{j}\left(w_{ji}, z_{ji}\right)\right) + \frac{1}{K}\sum_{j=1}^{K}\left(\mathrm{E}_{j}\left(w_{ji}\right)\mathrm{E}_{j}\left(z_{ji}\right)\right).\]

Using the notation \(\bar{w}_{j} = E_{j}(w_{ji})\) and
\(\bar{z}_{j} = E_{j}(z_{ji})\) to indicate the expectation in group
\(j\),

\[\frac{\Omega}{K n} = \left(\frac{1}{K}\sum_{j=1}^{K}\mathrm{Cov}_{j}\left(w_{ji}, z_{ji}\right)\right) + \mathrm{E}\left(\bar{w}_{j} \bar{z}_{j} \right).\]

We can also rewrite the first term on the right-hand side as an
expectation,

\[\frac{\Omega}{K n} = \mathrm{E}\left(\mathrm{Cov}_{j}\left(w_{j}, z_{j}\right)\right) + \mathrm{E}\left(\bar{w}_{j} \bar{z}_{j} \right).\]

We can rearrange the second term on the right-hand side
(\(\bar{\bar{w}}\) indicates grand mean over all groups),

\[\frac{\Omega}{K n} = \mathrm{E}\left(\mathrm{Cov}_{j}\left(w_{j}, z_{j}\right)\right) +  \mathrm{Cov}\left(\bar{w}_{j}, \bar{z}_{j} \right) + \bar{\bar{w}}\bar{\bar{z}}.\]

As in the main text, note that \(Kn\bar{\bar{w}}\) accounts for
differences in total population size from \(t\) to \(t+1\), with
\(\bar{\bar{w}}\) being mean fitness across all groups. We can therefore
set \(\Omega = Kn\bar{\bar{w}}\bar{\bar{z'}}\), so,

\[\frac{Kn\bar{\bar{w}}\bar{\bar{z'}}}{K n} - \bar{\bar{w}}\bar{\bar{z}} = \mathrm{E}\left(\mathrm{Cov}_{j}\left(w_{j}, z_{j}\right)\right) +  \mathrm{Cov}\left(\bar{w}_{j}, \bar{z}_{j} \right).\]

Because \(\Delta \bar{\bar{z}} = \bar{\bar{z'}} - \bar{\bar{z}}\),

\[\bar{\bar{w}}\Delta \bar{\bar{z}} = \mathrm{Cov}\left(\bar{w}_{j}, \bar{z}_{j} \right) + \mathrm{E}\left(\mathrm{Cov}_{j}\left(w_{j}, z_{j}\right)\right).
\tag{S2.2}
\]

Equation S2.2 therefore recovers the multi-level Price (1972) equation
(Lehtonen 2020) from a starting point of eco-evolutionary change in
different groups. Equation S2.2 can be found in Lehtonen (2016) B2.I,
who then derives a multi-level selection version of Hamilton's rule
predicting the evolution of altruism.

\section*{3. Integration of ecology and
evolution}

For simplicity, we now focus on showing an integration between ecology
and evolution using a population with no multi-level selection, and we
let \(\Delta z_{i} = 0\) (i.e., no transmission bias). As above in the
section on density-dependent population growth, we define
\(w_{i} = \beta_{i} - \delta_{i} + 1\) and use \(\alpha_{ij}\) to
represent the effect of individual \(j\) on the fitness of individual
\(i\). Our starting equation is therefore,

\[\Omega = \sum_{i=1}^{N}w_{i}\left(1 - \sum_{j=1}^{N}\alpha_{ij}\right)z_{i}.
\tag{S3.1}
\]

We have already demonstrated that if we assume all individuals have the
same effect on a focal individual such that \(\alpha = \alpha_{ij}\) for
all \(i\) and \(j\) pairs, we can recover equation S1.4 when
\(z_{i} = 1\) and \(\Omega\) is therefore interpreted as the count of
entities,

\[N_{t+1} = \bar{w} N_{t} \left(1 - \alpha N_{t}\right).\]

We now start from S3.1 to derive \(\Delta \bar{z}\). The objective is to
use our definition of eco-evolutionary change to simultaneously recover
how interactions between individuals affect population change and
evolutionary change.

We start by dividing both sides of S3.1 by \(N\),

\[\frac{\Omega}{N} = \frac{1}{N}\sum_{i=1}^{N}w_{i}\left(1 - \sum_{j=1}^{N}\alpha_{ij}\right)z_{i}.
\tag{S3.2}
\]

We can express the right-hand side of eqn S3.2 as an expectation,

\[\frac{\Omega}{N} = \mathrm{E}\left(w_{i}\left(1 - \sum_{j=1}^{N}\alpha_{ij}\right)z_{i}\right).\]

We can rewrite the right-hand side in terms of covariances,

\[\frac{\Omega}{N} = \mathrm{Cov}\left(w_{i}\left(1 - \sum_{j=1}^{N}\alpha_{ij}\right), z_{i}  \right) + \mathrm{E}\left(w_{i}\left(1 - \sum_{j=1}^{N}\alpha_{ij}\right)\right)\mathrm{E}\left(z_{i}\right). 
\tag{S3.3}
\]

The expectations in the second term on the right-hand side of eqn S3.3
can be replaced with overbars to represent the mean,

\[\frac{\Omega}{N} = \mathrm{Cov}\left(w_{i}\left(1 - \sum_{j=1}^{N}\alpha_{ij}\right), z_{i}  \right) + \overline{w_{i}\left(1 - \sum_{j=1}^{N}\alpha_{ij}\right)}    \bar{z}_{i}. 
\tag{S3.4}
\]

In the main text, we noted that \(\Omega = N\bar{w}\bar{z}'\). Here mean
fitness incorporates individual interactions, therefore,

\[\Omega = N\overline{w_{i}\left(1 - \sum_{j=1}^{N}\alpha_{ij}\right)}\bar{z}'.\]

We can therefore rewrite S3.4,

\[\overline{w_{i}\left(1 - \sum_{j=1}^{N}\alpha_{ij}\right)}\bar{z}' -  \overline{w_{i}\left(1 - \sum_{j=1}^{N}\alpha_{ij}\right)}    \bar{z}_{i} = \mathrm{Cov}\left(w_{i}\left(1 - \sum_{j=1}^{N}\alpha_{ij}\right), z_{i}  \right). \]

Noting again \(\Delta \bar{z} = \bar{z}' - \bar{z}\),

\[\overline{w_{i}\left(1 - \sum_{j=1}^{N}\alpha_{ij}\right)}\Delta\bar{z} = \mathrm{Cov}\left(w_{i}\left(1 - \sum_{j=1}^{N}\alpha_{ij}\right), z_{i}  \right). 
\tag{S3.5}
\]

We can rewrite the right-hand side of S3.5,

\[\overline{w_{i}\left(1 - \sum_{j=1}^{N}\alpha_{ij}\right)}\Delta\bar{z} = \mathrm{Cov}\left(w_{i} - w_{i}\sum_{j=1}^{N}\alpha_{ij}, z_{i}  \right). 
\tag{S3.6}
\]

The covariance term in S3.6 can be split without any additional
assumptions,

\[\overline{w_{i}\left(1 - \sum_{j=1}^{N}\alpha_{ij}\right)}\Delta\bar{z} = \mathrm{Cov}\left(w_{i}, z_{i}  \right) - \mathrm{Cov}\left(w_{i}\sum_{j=1}^{N}\alpha_{ij}, z_{i}  \right). 
\tag{S3.7}
\]

If we are able to further assume that \(w_{i}\) and the summation over
\(\alpha_{ij}\) are independent, then we could rewrite eqn S3.7,

\[\overline{w_{i}\left(1 - \sum_{j=1}^{N}\alpha_{ij}\right)}\Delta\bar{z} = \mathrm{Cov}\left(w_{i}, z_{i}  \right) - \mathrm{Cov}\left(\sum_{j=1}^{N}\alpha_{ij}, z_{i}  \right)\bar{w_{i}}. 
\tag{S3.8}
\]

Partitioning fitness into different components with the Price equation
is commonplace. But this derivation highlights, e.g., the ecological and
evolutionary relationship between nonsocial and social components of
fitness, and population size. For example, the second term on the
right-hand side of S3.8 shows the covariance between the sum of social
interactions and a trait. When traits covary with the interaction
between sociality and fitness, they will have a stronger effect on trait
change.

\section*{4. Expanded derivation of multi-level selection}

Here we expand on what we did in this supporting information section 2,
deriving multi-level evolutionary change without making the assumption
that \(\Delta z_{ji} = 0\). Our objective here is slightly different
from simply recovering multi-level selection. Here we suggest a way to
partition mean trait change in a community into components reflecting
biological processes occurring at the species and individual level.

To make this example more instructive, let \(z_{ji}\) be the mass of
individual \(i\) in species \(j\), and suppose that the focus concerns
the total biomass in the community. Let there be \(K\) species in this
community and \(N_{j}\) individuals per species \(j\). We can recover
the total biomass \(\Omega\) of the community at time \(t\) by summing
up the masses of all individuals \(i\) for each population of each
species \(j\),

\[\Omega_{t} = \sum_{j}^{K}\sum_{i}^{N_{j}}z_{ji}.
\tag{S4.1}
\]

In eqn S4.1, total biomass at time \(t + 1\) will be determined by the
absolute fitness of each individual
\(w_{ji} = \beta_{ji} - \delta_{ji} + 1\), where \(\beta_{ji}\) defines
the number of births from \(t\) to \(t+1\) attributable to individual
\(i\) in species \(j\), and \(\delta_{ji}\) is indicates individual
death (\(\delta_{ji} = 1\)) or survival (\(\delta_{ji} = 0\)). Changes
in individual mass \(z_{ji}\) are defined by \(\Delta z_{ji}\).
Consequently, total biomass at \(t+1\) equals,

\[\Omega_{t + 1} = \sum_{j}^{K}\sum_{i}^{N_{j}}w_{ji}\left(z_{ji} + \Delta z_{ji}\right).
\tag{S4.2}
\]

Again, note that if we instead define \(z_{ji}\) as the individual unit
\(z_{ji} = 1\) (i.e., an individual \(i\) contributes a count of 1 to
population \(j\)) and \(\Delta z_{ji} = 0\) (this contribution quantity
cannot change), then we recover total community size at \(t+1\) because
we are summing up all living individuals and their descendants. As in
section 2 of this supporting information, for simplicity, assume that
each species \(i\) has the same number of species, \(N_{j} = n\) for all
\(j\). Note that following from eqn S4.2,

\[\Omega_{t + 1} = \sum_{j}^{K}\sum_{i}^{n}\left(w_{ji}z_{ji} + w_{ji}\Delta z_{ji}\right).\]

We can divide both sides of the equation by \(Kn\),

\[\frac{\Omega_{t + 1}}{Kn} = \left(\frac{1}{K}\right)\left(\frac{1}{n}\right)\sum_{j}^{K}\sum_{i}^{n}\left(w_{ji}z_{ji} + w_{ji}\Delta z_{ji}\right).
\tag{S4.3}
\]

We can rewrite eqn S4.3,

\[\frac{\Omega_{t + 1}}{Kn} = \frac{1}{K}\sum_{j}^{K}\left(\frac{1}{n}\sum_{i}^{n}\left(w_{ji}z_{ji} + w_{ji}\Delta z_{ji}\right)\right).\]

We can separate the inner summation,

\[\frac{\Omega_{t + 1}}{Kn} = \frac{1}{K}\sum_{j}^{K}\left(\frac{1}{n}\sum_{i}^{n}\left(w_{ji}z_{ji}\right) + \frac{1}{n}\sum_{i}^{n}\left(w_{ji}\Delta z_{ji}\right)\right).\]

These inner summations can be rewritten as expectations. For this, we
will drop the individual subscripts \(i\),

\[\frac{\Omega_{t + 1}}{Kn} = \frac{1}{K}\sum_{j}^{K}\left(\mathrm{E}\left(w_{j}z_{j}\right) + \mathrm{E}\left(w_{j}\Delta z_{j}\right)\right).
\tag{7}
\]

Because
\(\mathrm{Cov}(X, Y) = \mathrm{E}(XY) - \mathrm{E}(X)\mathrm{E}(Y)\),

\[\frac{\Omega_{t + 1}}{Kn} = \frac{1}{K}\sum_{j}^{K}\left(\mathrm{Cov}\left(w_{j},z_{j}\right) + \mathrm{E}\left(w_{j}\right)\mathrm{E}\left(z_{j}\right) + \mathrm{Cov}\left(w_{j},\Delta z_{j}\right) + \mathrm{E}\left(w_{j}\right)\mathrm{E}\left(\Delta z_{j}\right)\right).\]

We can separate the summation across \(K\),

\[\frac{\Omega_{t + 1}}{Kn} = \frac{1}{K}\sum_{j}^{K}\left(\mathrm{Cov}\left(w_{j},z_{j}\right)\right) + \frac{1}{K}\sum_{j}^{K}\left(\mathrm{E}\left(w_{j}\right)\mathrm{E}\left(z_{j}\right)\right) + \frac{1}{K}\sum_{j}^{K}\left(\mathrm{Cov}\left(w_{j},\Delta z_{j}\right)\right) + \frac{1}{K}\sum_{j}^{K}\left(\mathrm{E}\left(w_{j}\right)\mathrm{E}\left(\Delta z_{j}\right)\right).
\tag{S4.4}
\]

We can rewrite terms 2 and 4 on the right-hand side of eqn S4.4 as
expectations. To make this easier to follow, we will use the notation
\(\bar{w_{j}} = \mathrm{E}(w_{j})\) and
\(\bar{z_{j}} = \mathrm{E}(z_{j})\),

\[\frac{\Omega_{t + 1}}{Kn} = \frac{1}{K}\sum_{j}^{K}\left(\mathrm{Cov}\left(w_{j},z_{j}\right)\right) + \mathrm{E}\left(\bar{w_{j}}\bar{z_{j}}\right) + \frac{1}{K}\sum_{j}^{K}\left(\mathrm{Cov}\left(w_{j},\Delta z_{j}\right)\right) + \mathrm{E}\left(\bar{w_{j}}\overline{\Delta z_{j}}\right).
\tag{S4.5}
\]

Note that the overbar in S4.5 spans \(\overline{\Delta z_{j}}\) in the
last term to indicate that this is the expected change in \(z_{j}\), and
\emph{not} the change in expected \(z_{j}\) (which might be different).
From here, we can also get expectations over \(K\) for the covariance
terms,

\[\frac{\Omega_{t + 1}}{Kn} = \mathrm{E}\left(\mathrm{Cov}\left(w_{j},z_{j}\right)\right) + \mathrm{E}\left(\bar{w_{j}}\bar{z_{j}}\right) + \mathrm{E}\left(\mathrm{Cov}\left(w_{j},\Delta z_{j}\right)\right) + \mathrm{E}\left(\bar{w_{j}}\overline{\Delta z_{j}}\right).
\tag{S4.6}
\]

Using the definition of covariance, we can expand terms 2 and 4 on the
right-hand side of eqn S4.6,

\[\frac{\Omega_{t + 1}}{Kn} = \mathrm{E}\left(\mathrm{Cov}\left(w_{j},z_{j}\right)\right) + \mathrm{Cov}\left(\bar{w_{j}}, \bar{z_{j}}\right) + \mathrm{E}\left(\bar{w_{j}}\right)\mathrm{E}\left(\bar{z_{j}}\right) + \mathrm{E}\left(\mathrm{Cov}\left(w_{j},\Delta z_{j}\right)\right) + \mathrm{Cov}\left(\bar{w_{j}}, \overline{\Delta z_{j}}\right) + \mathrm{E}\left(\bar{w_{j}}\right)\mathrm{E}\left(\overline{\Delta z_{j}}\right).\]

Again, to make this easier to follow, we will use the notation
\(\bar{\bar{w}} = \mathrm{E}(\bar{w}_{j})\) and
\(\bar{\bar{z}} = \mathrm{E}(\bar{z}_{j})\),

\[\frac{\Omega_{t + 1}}{Kn} = \mathrm{E}\left(\mathrm{Cov}\left(w_{j},z_{j}\right)\right) + \mathrm{Cov}\left(\bar{w_{j}}, \bar{z_{j}}\right) + \bar{\bar{w}}\bar{\bar{z}} + \mathrm{E}\left(\mathrm{Cov}\left(w_{j},\Delta z_{j}\right)\right) + \mathrm{Cov}\left(\bar{w_{j}}, \overline{\Delta z_{j}}\right) + \bar{\bar{w}}\overline{\overline{\Delta z}}.
\tag{S4.7}
\]

It is now a good time to define \(\Omega_{t+1}\) in terms of the sum
amount of mass, i.e., the total biomass in a community at \(t + 1\). As
in section 2, this will be the product of the grand mean fitness
(\(\bar{\bar{w}}\)) and the grand mean trait of all individuals in the
community at \(t+1\) (\(\bar{\bar{z}}'\)). Note that
\(\bar{\bar{w}}\bar{\bar{z}}'\) is the grand mean fitness at \(t\) times
the grand mean trait of descendants at \(t+1\), which we multiply by the
total number of individuals in the community \(Kn\) to get
\(\Omega_{t+1} = Kn\bar{\bar{w}}\bar{\bar{z}}'\). We can therefore
rewrite eqn S4.7,

\[\bar{\bar{w}}\bar{\bar{z}}' = \mathrm{E}\left(\mathrm{Cov}\left(w_{j},z_{j}\right)\right) + \mathrm{Cov}\left(\bar{w_{j}}, \bar{z_{j}}\right) + \bar{\bar{w}}\bar{\bar{z}} + \mathrm{E}\left(\mathrm{Cov}\left(w_{j},\Delta z_{j}\right)\right) + \mathrm{Cov}\left(\bar{w_{j}}, \overline{\Delta z_{j}}\right) + \bar{\bar{w}}\overline{\overline{\Delta z}}.
\tag{S4.8}
\]

Since \(\Delta \bar{\bar{z}} = \bar{\bar{z}}' - \bar{\bar{z}}\), i.e.,
the change in the grand mean of the trait (mass in this case) is the
grand mean at \(t+1\) minus the grand mean at \(t\), we can subtract
\(\bar{\bar{w}}\bar{\bar{z}}\) from both sides of eqn S4.8 and
rearrange,

\[\bar{\bar{w}}\Delta\bar{\bar{z}} = \mathrm{E}\left(\mathrm{Cov}\left(w_{j},z_{j}\right)\right) + \mathrm{Cov}\left(\bar{w_{j}}, \bar{z_{j}}\right) +  \mathrm{E}\left(\mathrm{Cov}\left(w_{j},\Delta z_{j}\right)\right) + \mathrm{Cov}\left(\bar{w_{j}}, \overline{\Delta z_{j}}\right) + \bar{\bar{w}}\overline{\overline{\Delta z}}.
\tag{S4.9}
\]

Finally, we rearrange the terms on the right-hand side of S4.9,

\[\bar{\bar{w}}\Delta\bar{\bar{z}} = \mathrm{Cov}\left(\bar{w_{j}}, \bar{z_{j}}\right) + \mathrm{E}\left(\mathrm{Cov}\left(w_{j},z_{j}\right)\right) + \mathrm{Cov}\left(\bar{w_{j}}, \overline{\Delta z_{j}}\right) +   \mathrm{E}\left(\mathrm{Cov}\left(w_{j},\Delta z_{j}\right)\right) + \bar{\bar{w}}\overline{\overline{\Delta z}}.
\tag{S4.10}
\]

In S4.10, we have a Price-like equation that describes the mean change
in trait across all individuals and species. Note that the first two
terms on the right-hand side are the same as found in Lehtonen (2016)
and section 2 of this supporting information, which therefore recovers
multi-level selection.

More investigation is needed, but an interesting approach might be to
consider how the different terms in S4.10 could correspond to ecological
and evolutionary processes. For example, we might be able to interpret
the first term \(\mathrm{Cov}\left(\bar{w_{j}}, \bar{z_{j}}\right)\) as
the effects of selection and drift at the level of species, i.e.,
potentially differences in mean fitness (growth rates) of species
attributable to differences in a specific trait such as body mass. The
second term
\(\mathrm{E}\left(\mathrm{Cov}\left(w_{j},z_{j}\right)\right)\) could be
attributed to selection and genetic drift on body mass within species.
The third term
\(\mathrm{Cov}\left(\bar{w_{j}}, \overline{\Delta z_{j}}\right)\) might
reflect changes in species fitness that are attributable to expected
changes in the mean species trait. This could include demographic change
in a population. For example, the average individual in a species might
increase in mass due to the changing demographics of a population, so
the expected change in total biomass of a species
\(\overline{\Delta z_{j}}\) might be correlated the population growth
rate of the species. The fourth term is the covariance between the
change in body mass and fitness,
\(\mathrm{E}\left(\mathrm{Cov}\left(w_{j},\Delta z_{j}\right)\right)\).
Conceptually, do individuals that experience a change in body mass
increase (or decrease) their fitness? This could potentially reflect
mutation, individual development, or plasticity. Finally, we have the
term \(\bar{\bar{w}}\overline{\overline{\Delta z}}\), which is the
product between the mean fitness across all individuals in the community
and the expected change in a trait for any individual in the community.

Note that some biological processes might be represented by a shift
between terms. For example, speciation will likely be represented by
some quantity shifting from the second term to the first term. This is
because variation within species has, by definition, been lost (a subset
of individuals in the population now belong to a new species). But we
also have more variation among species (because there are more species
with potentially different mean traits). Overall, we believe that this
framework introduces new avenues for integrating ecological and
evolutionary theory from first principles.

\section*{Supporting Information References}

\hangindent=0.5cm
Lehtonen, J. 2016.
\href{https://doi.org/10.1016/j.tree.2016.07.006}{{Multilevel selection
in kin selection language}}. Trends in Ecology and Evolution 31:752--762.

\hangindent=0.5cm
---------. 2020. {``The Price Equation and the Unity of Social Evolution
Theory.''} \emph{Philosophical Transactions of the Royal Society B:
Biological Sciences} 375:20190362.
\url{https://doi.org/10.1098/rstb.2019.0362}.

\hangindent=0.5cm
Lion, S. 2018. \href{https://doi.org/10.1086/694865}{{Theoretical
approaches in evolutionary ecology: environmental feedback as a unifying
perspective}}. American Naturalist 191:21--44.

\hangindent=0.5cm
Lion, S., A. Sasaki, and M. Boots. 2023. {``Extending
Eco-Evolutionary Theory with Oligomorphic Dynamics.''} \emph{Ecology
Letters} 26: S22--46.
\url{https://doi.org/10.1111/ele.14183}.

\hangindent=0.5cm
Price, G. R. 1972. {``{Extension of covariance selection
mathematics}.''} \emph{Annals of Human Genetics} 35: 485--90.
\url{https://doi.org/10.1111/j.1469-1809.1957.tb01874.x}.

\end{document}